# The effect of shape anisotropy on the spectroscopic characterization of the magneto-optical activity of nanostructures

Guan-Xiang. Du*, Shin Saito, and Migaku Takahashi
Department of Electronic Engineering, Graduate School of Engineering, Tohoku University, 6-6-05 Aoba, Aramaki, Aoba-ku, Sendai 980-8579, Japan
*guanxiang.du@unibas.ch

**Abstract**: How to measure magnetic field induced magneto-optical (MO) activity of nonmagnetic elliptical plasmonic nanodisks which rest on a dielectric substrate remains a challenge since the substrate contribute most of the overall MO which varies with light polarization with respect to the orientation of the nanodisks. Here we present a spectroscopic characterization. We find that only when light is incident from the nanostructures' side with polarization aligned with one of the two symmetry axes, one can subtract the MO contribution from the substrate by an amount equal to that of a bare one. By a detailed polarizing transmittance measurement we determine the orientation of the two symmetry axes of the nanodisks. Light polarization is then aligned along the axes, enabling measurement of the intrinsic MO activity of gold nanodisks, which is the overall MO activity subtracted by that of a bare glass substrate. The narrow line widths of the plasmonic resonance features in the MO spectra imply a potential application in refractive index sensing.

**PACS Numbers**: 73.20.Mf, 78.20.Ls, 79.60.Jv

**Keywords**: localized surface plasmon resonance (LSPR), plasmonics, magneto-optic effect.

## Introduction

Plasmonics hold promise for generating, guiding, manipulating and detecting light signals down to nanometer scale optical components, which is far smaller than the diffraction limit [1- 4]. However, most plasmonic building blocks are passive, i.e. their functions cannot be programmed by an external field once they are fabricated on chips. Magnetoplasmonics has emerged in recent years as an active candidate [5 - 7], allowing us to manipulate the light polarization state by external magnetic field based on magneto-optical (MO) effect. One approach is to incorporate MO active materials such as Co, Ni, Fe or rare earth elements into noble metals [8 - 10]. The ferromagnetic materials provide sizable tunability on the polarization under moderate magnetic field. However, they are dispersive associated with a large imaginary part in the dielectric constant [11]. Another approach is based only on noble metals like Au, Ag and Cu. The MO activity of bulk gold originates from the Lorentz force, where light drives electrons to oscillate at the optical frequency in a magnetic field, resulting in a rotation of the light polarization. This effect is normally 3 orders of magnitude smaller than that of ferromagnetic substitutes. However, it can be significantly enhanced at plasmonic resonance [12] in gold nanostructures. A technical issue in characterizing the intrinsic MO activity of the gold nanostructures is how to deal with the contribution from the substrate on which gold nanostructures sit on. For thin films, the measurement of MO activity is based on the fact that the contribution from substrate is independent of the polarization angle of incident linear polarized light and can be subtracted from the total MO activity. This is consistent with the uniaxial symmetry of thin films around the magnetization axis. However for anisotropic gold nanostructures, the uniaxial symmetry breaks and the polarization state of light is geometrically changed by the birefringence of nanostructures. Analysis of the Jones vectors suggests that both the intrinsic MO activity of substrate and the birefringence of nanostructures should be taken into account to understand why the nominal (non-intrinsic) contribution from substrate is dependent on the polarization [13].

This work is organized as follows. First we show how we formulate the nominal MO activity spectra based on both the intrinsic MO activity of glass substrate and the birefringence of gold elliptical nanodisks. We determine that when light is incident from the nanodisks side and the polarization aligned along the major or minor axis, the nominal MO activity of substrate is identical to the intrinsic one if neglecting the MO from gold nanodisks. However, it is difficult to separate the intrinsic MO activity from the total one since the nominal MO activity of substrate is very sensitive to the incident polarization. Next, we measured the intrinsic MO activity of nearly circular gold nanodisks where the polarization dependence is less sensitive than in case of elliptical ones.

## Experiments, results and discussion

The elliptical and circular (not perfect) gold nanodisk arrays were fabricated by electron beam lithography (EBL) together with argon ion milling. The gold film was deposited on a 0.55 mm thick borosilicate glass substrate (D263, Schott) by sputtering a 2 nm Ti as adhesive layer. The thickness for circular and elliptical nanodisks was 40 and 80 nm, respectively. Both nanodisks were arranged in a square array with grating constant of 250 nm. The electromagnetic coupling between nanodisks, termed as Bragg diffraction in periodic array has been extensively discussed [14,8]. However, with the small grating constant, this effect is absent in the measurement wavelength range from 530 to 960 nm. A resist of 150 nm thick was spin-coated to cover the patterned nanostructures with refractive index matched to the glass substrate, creating an isotropic dielectric environment for gold nanostructures. The lateral dimension of the nanostructures was measured by scanning electron microscope (SEM).

We have developed a wavelength-parallel spectroscopic system to characterize the optic transmittance and the Faraday rotation or ellipticity of the plasmonic nanostructures by microscope that was described elsewhere [15]. As illustrated in Fig. 1 (a), the light source is a halogen lamp. The polarization change, or the Faraday rotation and ellipticity, is measured by a pair of polarizers (P and A) detuned from the crossed Nicol configuration with the polarizer angle $\varphi$ variable and the analyzer fixed at -45°. The Jones vectors are shown in Fig. 1 (a). For isotropic materials, the Faraday rotation is given by ($\varphi = 0$)

$$\theta_F(\lambda_i,H) = -\tfrac{1}{2}\frac{I(\lambda_i,H)-I(\lambda_i,0)}{I(\lambda_i,0)} = -\tfrac{1}{2}\frac{I(\lambda_i,H)-I(\lambda_i,-H)}{I(\lambda_i,H)+I(\lambda_i,-H)} \quad (1)$$

where $I(\lambda_i,H)$ denotes the intensity of the pixel in the CCD array with wavelength of $\lambda_i$ at magnetic field $H$. To measure the Faraday ellipticity, an achromatic quarter wave plate (QW) is inserted and the same equation applies. A series of lenses (L1 – L4) are used for collimation and focus of white light beam.

Since no modulation is used in this system, The MO spectrum loops are obtained by recording the light intensity in the CCD linear array at user-defined magnetic fields. The noise level is dominated by the intensity instability of halogen lamp and discussed in Ref. 15. An angle resolution of 0.004° was demonstrated, at the same order of that by polarization modulation method (0.001°) [16, 17].

First, the birefringence originating from the biaxial shape of elliptical nanodisks is characterized and its effect on the measurement of MO activity is formulated. The transmission tensor of the elliptical nanodisks is characterized by a Jones matrix $\begin{pmatrix} t_l & 0 \\ 0 & t_s \end{pmatrix}$. Normally, the Fresnel transmission coefficients $t_l$ and $t_s$ are complex values, $t_l \equiv |t_l| e^{i\delta_l}$ and $t_s \equiv |t_s| e^{i\delta_s}$. $|t_l|^2$ and $|t_s|^2$ correspond to the transmittance for light polarizing along the long and short axis, respectively. We applied the method described in Ref. 13 to determine the phase difference $\Delta \equiv \delta_l - \delta_s$ as follows. Difference is that here we use white light source, rather than laser diodes. First, the short axis of the elliptical nanodisk was aligned along horizontal ($x$) axis. This was done by placing the sample between a pair of crossed polarizers and rotating the sample to minimize the transmitted light since in this case the ellipticity of light remains unchanged even in the presence of shaped nanodisks. Then the analyzer angle was fixed at −45° while the polarizer angle $\varphi$ varied between 0° to −90°. The transmittance, which is the ratio of the light intensity passing through the nanodisks to that through bare glass substrate, is given by

$$T_{45} = \left(|t_s|^2 \cos^2\varphi + |t_l|^2 \sin^2\varphi - |t_s t_l| \sin 2\varphi \cos\Delta\right) \Big/ (1 - \sin 2\varphi) \qquad (2)$$

For $\varphi$ = 0° and −90°, $T$ equals $|t_s|^2$ and $|t_l|^2$, respectively.

The gold elliptical nanodisks have lateral dimension of 77 × 174 (nm) as shown in Fig. 1 (b). The transmittance spectra $T_{45}$ at different polarizer angles $\varphi$ are shown in Fig. 2 (a). The dips at 618 ($\varphi$ = 0°) and 759 nm ($\varphi$ = −90°) correspond to the transverse and longitudinal plasmon mode, which are excited when light polarization is along short and long axis of the elliptical disks, respectively. From Eq. (2), the wavelength dependent phase difference $\Delta$ could be deduced, associated with its complex birefringence. The results are shown in Fig. 2 (b). At around 680 nm, the phase difference takes maxima, corresponding to a minimum in cos($\Delta$). Note that at this wavelength, the transmittance along the short and long axis of the elliptical nanodisk equals, i.e. $|t_l| = |t_s|$, as indicated in Fig. 2 (a) and (b). This is a general feature observed in our shaped nanodisks.

Since the amplitude of Faraday rotation angle of gold nanodisks is much weaker than that of the glass substrate, which has millimeter thickness, compared to 40 or 80 nm for gold nanodisks [12], we neglect the MO activity of the elliptical nanodisk. Following Ref. 13, Let's consider that the light passes through the gold nanodisks and glass substrate in a sequence model, where the nanodisks orientated at arbitrary angle $\varphi$ changes the polarization state of incident light due to different refractive index for polarization along its long and short axis. For top incidence, i.e., linear polarized light first arrives at the nanodisks and then propagates through the glass substrate, experiencing a Faraday rotation of $\theta_g$ in glass, which is proportional to applied magnetic field $H$. The nominal (nonintrinsic) Faraday rotation following the definition of Eq. (1) is thus given by

$$\frac{I^+ - I^-}{2\left(I^+ + I^-\right)} = -\theta_g \frac{|t_s|^2 \cos^2\varphi - |t_l|^2 \sin^2\varphi}{|t_s|^2 \cos^2\varphi + |t_l|^2 \sin^2\varphi - |t_s t_l| \sin 2\varphi \cos\Delta} \quad (3)$$

where $I^+$ and $I^-$ denote the light intensity at magnetic field $H$ and $-H$, respectively. For bottom incidence, light first passes through glass substrate experiencing a rotation of $\theta_g$ and then transmits through the nanodisks,

$$\frac{I^+ - I^-}{2\left(I^+ + I^-\right)} = -\theta_g \frac{\frac{1}{2}\left(|t_s|^2 - |t_l|^2\right)\sin 2\varphi + |t_s t_l| \cos 2\varphi \cos\Delta}{|t_s|^2 \cos^2\varphi + |t_l|^2 \sin^2\varphi - |t_s t_l| \sin 2\varphi \cos\Delta} \quad (4)$$

Both equations indicate that the nominal Faraday rotation measured in this way, differs from the intrinsic one ($\theta_g$) by a factor related to the complex birefringence of the elliptical nanostructures. Note that the nominal MO activity depends on not only the ratio of the transmission coefficients $|t_l|/|t_s|$, but also the phase difference $\Delta$.

Fig. 2 (c) and (e) show the nominal Faraday rotation measured at top and bottom incidence, respectively. The calculation results shown in Fig. 2 (d) and (f) are based on Eq. (3) and (4), using the intrinsic Faraday rotation of glass $\theta_g$ measured at $\varphi = 0°$ and the birefringence data shown in Fig. 2 (a) and (b). Note that in the case of bottom incidence at around 780 nm and light applied along long axis, the nominal Faraday rotation increased to 0.3 deg/Tesla, while the intrinsic value of glass substrate is 0.08 deg/Tesla at this wavelength, which is the tailoring effect of shaped nanodisk on the glass MO as first described in Ref. 13. We can also see that, by set $\varphi$ to 0 or -90°, the MO activity measured at these two polarizer angle with top incident light is equal to that of bare glass substrate, i.e. its intrinsic value. This tells that one can measure the intrinsic MO activity of gold nanodisks at top incidence with polarization aligned along the symmetry axes of the nanostructures, by subtracting the MO activity of a bare glass substrate. However, to obtain the relatively small intrinsic MO activity of gold nanostructures, the sensitivity on polarization angle should be further minimized since a small misalignment of the polarizer may result a significant deviation of the MO contribution from substrate. Therefore, we fabricated circular gold nanodisks to measure its intrinsic MO activity, with care paid on minimizing the geometrically introduced error.

Indeed, we will show below that even non-obvious distortion of circular gold nanodisk results in significant birefringence. Due to astigmation in the adjustment of electron beam optics, lithography defined circular structures are usually distorted but not obvious even from SEM images, which serves as a common tool to image nanostructures. Since the distortion is small, as a first order approximation, these nanodisks can be treated to elliptical having long axis along a specific direction in the plane of disk array. The distortion was evaluated by polarizing transmittance spectra, especially sensitive to light polarization at plasmonic resonance. The sample was fixed and located between the polarizer and analyzer. The transmitted light intensity was obtained by rotating the polarizer pair while keeping them in crossed Nicol configuration. As shown in the inset of Fig. 3 (a), $\varphi_0$ is the angle of the long axis of the distorted nanodisk, the transmission axis of the polarizer is $\varphi_P$ with respect to the horizontal axis, the transmitted light observes the relationship of $I = I_0 |t_s - t_l|^2 \sin^2(\varphi_P - \varphi_0)\cos^2(\varphi_P - \varphi_0)$ where $I_0$ is the intensity of light source. From this equation, it is found that when the transmitted light is at minimum, i.e. $\varphi_P = \varphi_0$, the polarizer transmission axis is aligned with the long or short axis of the nanodisks.

Fig. 3 (a) shows the polarizing transmittance at three typical wavelengths. For all wavelengths, the minimum light intensity occurs at $\varphi_P$ = -30°, which is consistent with the size and orientation fitting of the SEM images as shown in Fig. 3 (b), The nonzero minimum intensity is caused by the depolarization effect in the objective lenses, observable though small. The unsymmetrical angle dependence is due to the polarization dependence of the grating efficiency of the spectrometer. Fig. 3 (c) shows its polarizing transmittance along the short and long axis. It is clearly seen though smaller compared to results in Fig. 2 (a). Fig. 3 (d) shows the transmittance ratio along the long axis to the short one and the phase difference deduced based on Eq. (2). Maximum phase difference locates at the wavelength where transmittance along long and short axis equals, consistent with the results in Fig. 2 (b).

To measure the intrinsic MO activity of circular gold nanodisks, light was top incident and polarization parallel to the long or short axis of nanodisks. Under this condition, the contribution from glass substrate is identical to that of a bare glass substrate. In Fig. 3 (e), the black line shows the Faraday rotation of bare glass substrate by slightly shifting laser spot to the region without nanodisks, the red line shows that measured on gold nanodisks with polarization aligned with the long axis, their difference is attributed to the intrinsic MO activity of gold nanodisks as summarized in Fig. 3 (f). To show how small difference in light polarization brings observable deviation, the nominal Faraday rotation measured on gold nanodisks with polarization orientated ±8 degree to the long axis are shown in Fig. 3 (e) by blue and green lines, respectively. At these angles, we also calculated the contributions from the glass substrate, shown by dashed lines, respectively. The difference between the calculated value and experimental value is MO activity of gold nanodisk at these polarizer angles. To be noted, MO of gold disks is not taken into account in Eq. (3), neither in Eq. (4). The dip position of the MO activity in Fig. 3 (f) corresponds to that of the transmittance suggesting plasmon enhancement. It is observed that (1) the sign of gold MO activity is opposite to that of the glass substrate, and (2) the line width of MO is smaller than that of the transmittance. This is a direct result of the polarizability tensor of gold nanodisks, of which the diagonal component is inversely proportional to $\varepsilon_{xx}+2\varepsilon_d$ while the off-diagonal one to $(\varepsilon_{xx}+2\varepsilon_d)^2$, where $\varepsilon_{xx}$ and $\varepsilon_d$ are the dielectric constant of gold and its surrounding dielectrics, respectively [11, 12]. The narrowed line width in the MO activity suggests applications requiring a higher quality factor.

## Conclusion

Spectroscopic MO activity of gold nanodisks have been presented. We have determined the formulism in the analysis of the optical and MO properties of plasmonic nanostructures seated on dielectric substrate. Emphasis is put on the anisotropy of nanostructures and how it correlates the Jones vectors of the incident and transmitted light. We analyzed the polarizing transmittance of gold circular nanodisks and revealed that their shape was elongated. The intrinsic MO activity of gold nanodisks was measured along the symmetry axis. Compared to the transmittance spectra, the MO spectra shows a much narrower line width, namely a higher quality factor which suggests characterizing the MO activity might be a promising method for refractive index sensing using plasmonic nanostructures [18-21].

## Acknowledgement

Support from the foreign postdoctoral program from JSPS (Grant-in-Aid No. 2109282) is acknowledged.

# Figures and captions

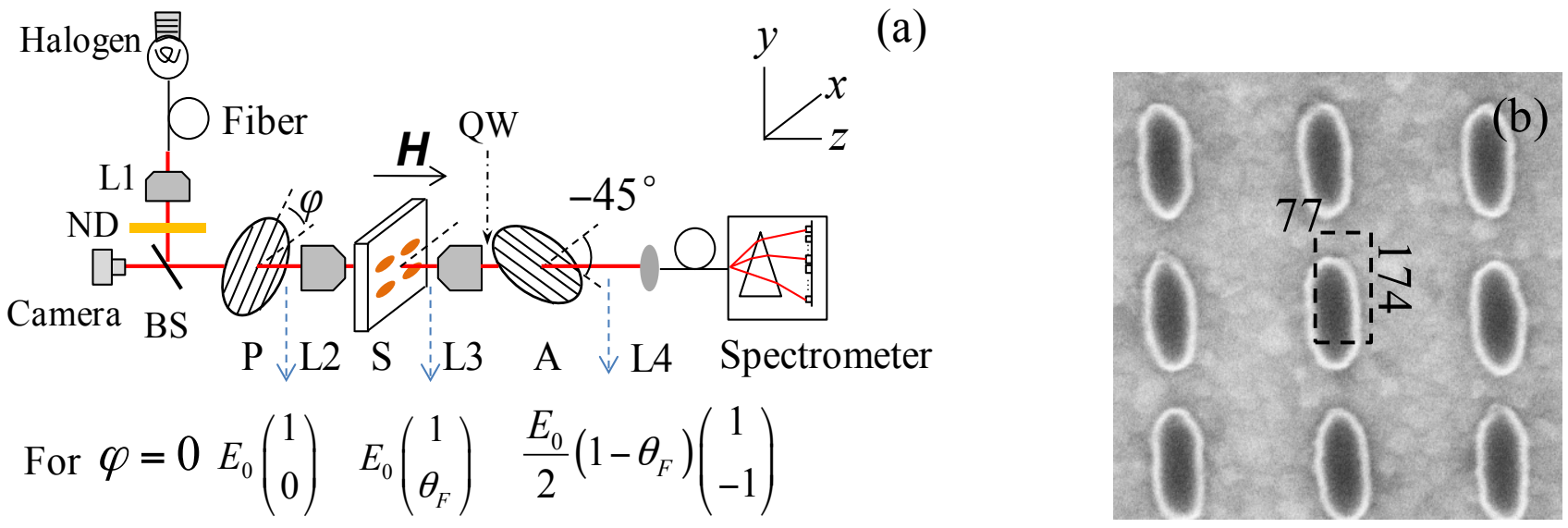

Fig. 1 (a) Sketch of the spectroscopic MO system. (b) SEM image of elliptical gold nanodisk array with lateral size of 77 × 174 nm. The grating constant is 250 nm.

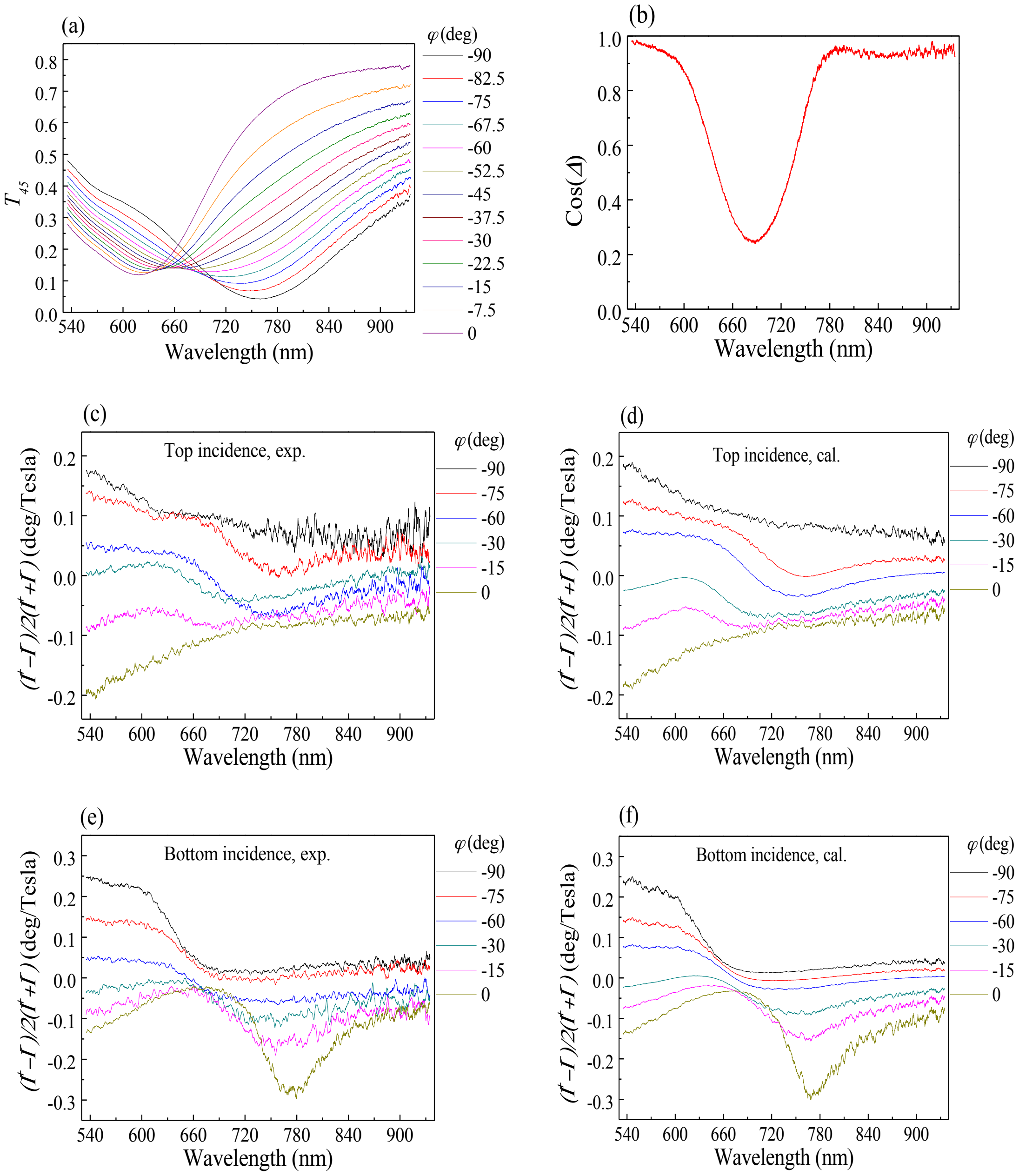

Fig. 2 Birefringence of elliptical gold nanodisks and its effect on the nominal Faraday rotation. (a) Transmittance spectra for various polarizer angles and the analyzer is fixed at −45°. Short axis of the nanorod is aligned along $x$-axis. (b) Phase difference cos($\Delta$) deduced from (a) based on Eq. (2). Nominal MO activity of the glass substrate was measured at various polarizer angles $\varphi$ for top (c) and bottom (e) incidence. Corresponding calculation results based on Eq. (3) and (4) are shown in (d) and (f), respectively.

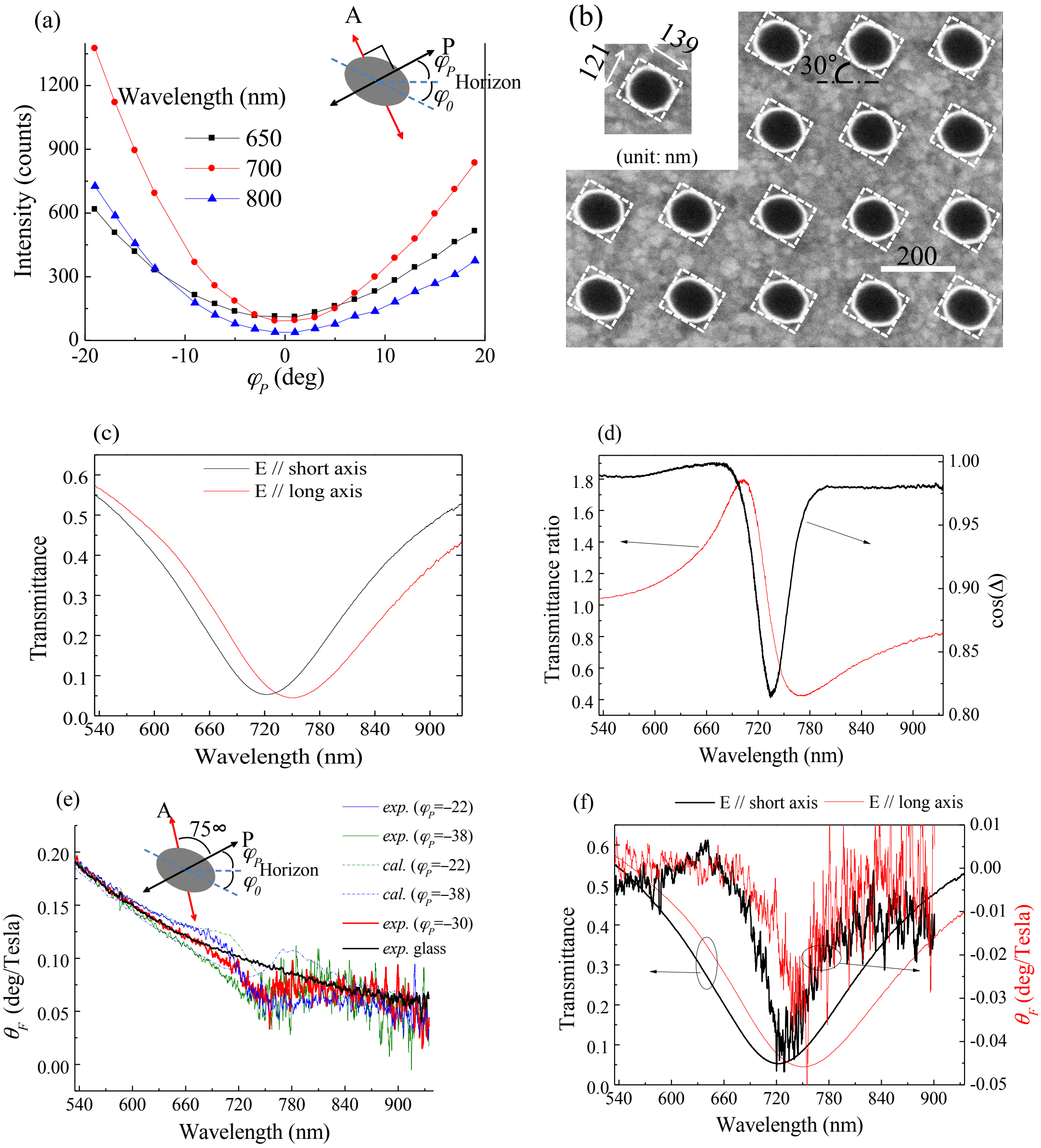

Fig. 3 Birefringence of the distorted circular gold nanodisks and its intrinsic MO activity. (a) Symmetry axis of the distorted nanodisks characterized by cross-polarizer method, minimum transmitted intensity indicates polarizer along symmetric axis. (b) Size fitting to the SEM images reveals circular nanodisks were elliptically distorted with long axis orientated at 30° with respect to the horizontal axis. (c) Transmittance measured along its long and short axis. (d) Transmittance ratio along the long to short axes and cosine of the phase difference of Fresnel transmission coefficient along the long and short axes. (e) MO activity of gold nanodisks measured along and deviated from the long axis. The modified MO activity of glass substrate by nanodisks was calculated. (f) Intrinsic MO activity of gold nanodisks and the corresponding transmittance measured along the long and short axis.